\documentstyle[prb,aps,twocolumn,floats,epsfig]{revtex}

\begin{document}
\draft
\wideabs{
\title{\bf Numerical atomic orbitals for linear-scaling calculations }
\author{Javier Junquera,$^1$ \'Oscar Paz,$^1$ Daniel S\'anchez-Portal,$^2$
and Emilio Artacho$^{3}$}
\address{
$^1$Departamento de F\'{\i}sica de la Materia Condensada, C-III,
Universidad Aut\'onoma, 28049 Madrid, Spain \\
$^2$Department of Physics and Materials Research Laboratory, 
University of Illinois, Urbana Illinois 61801, USA \\
$^3$Department of Earth Sciences, University of Cambridge, Downing Street,
Cambridge CB2 3EQ, UK}
\date{\today}
\maketitle

\begin{abstract}
  The performance of basis sets made of numerical atomic orbitals 
is explored in density-functional calculations of solids and molecules.
  With the aim of optimizing basis quality while maintaining strict
localization of the orbitals, as needed for linear-scaling calculations,
several schemes have been tried. 
  The best performance is obtained for the basis sets generated according
to a new scheme presented here, a flexibilization of previous proposals.
  The basis sets are tested versus converged plane-wave calculations
on a significant variety of systems, including covalent, ionic and metallic.
  Satisfactory convergence (deviations significantly smaller than
the accuracy of the underlying theory) is obtained for reasonably small
basis sizes, with a clear improvement over previous schemes.
  The transferability of the obtained basis sets is tested in several
cases and it is found to be satisfactory as well.
\end{abstract}

}


\section{Introduction}

  In order to make intelligent use of the increasing power of computers
for the first-principles simulation of ever larger and more complex
systems, it is important to develop and tune linear-scaling
methods, where the computational load scales only linearly with the 
number of atoms in the simulation cell.
  The present status of these methods and their applications
can be found in several reviews.\cite{Ordejon98,Goedecker}
  Essential for linear scaling is locality, and basis sets made 
of localized wavefunctions represent a very sensible basis choice.
  It is not only the scaling what matters, however, the prefactor being
also important for practical calculations. 
  The prefactor depends significantly on two aspects of the
basis: ($i$) the number of basis functions per atom, and
($ii$) the size of the localization regions of these functions.

  Atomic orbitals offer efficient basis sets since, even though their 
localization ranges are larger than those of some other 
methods,\cite{Hernandez} the number of basis functions needed is 
usually quite small.
  The price to pay for this efficiency is the lack of systematics for
convergence. Unlike with plane-wave\cite{Payne} 
or real-space-grid\cite{realgrid} related methods,
there is no unique way of increasing the size of the basis, and the
rate of convergence depends on the way the basis is enlarged.
  This fact poses no fundamental difficulties, it just means that some
effort is needed in the preparation of unbiased basis sets, in analogy to the
extra work required to prepare pseudopotentials to describe the effect of
core electrons.

  Maximum efficiency is achieved by choosing atomic orbitals
that allow convergence with small localization ranges 
{\it and} few orbitals. 
  It is a challenge again comparable to the one faced by the pseudopotential
community, where transferability {\it and} softness are sought.\cite{tm2}
  For atomic wavefunctions the optimization freedom is in the radial shape.
  Gaussian-type orbitals have been proposed for linear 
scaling,\cite{Head-Gordon,Scuseria,Hutter} connecting with the tradition
of quantum chemistry.\cite{Huzinaga,Poirier} 
  These bases are, however, quite rigid for the mentioned optimization, 
imposing either many gaussians or large localization ranges.

  Numerical atomic orbitals (NAOs) are more flexible in this respect.
  Different ideas have been proposed in the literature, originally within
tight-binding contexts concentrating on minimal (single-$\zeta$) bases. 
  They are obtained by finding the eigenfunctions of the isolated atoms 
confined within spherical potential wells of different 
shapes,\cite{Sankey,Porezag,Horsfield} or directly modifying 
the eigenfunctions of the free atoms.\cite{Kenny} 
  These schemes give strictly localized orbitals, i.e., orbitals that 
are strictly zero beyond given cutoff radii $r_c$.
  A first extension towards more complete basis sets was proposed using
the excited states of the confined atoms,\cite{Daniel-JPCM} but the
quite delocalized character of many excited states made this approach
inefficient unless very stringent confinement potentials were 
used.\cite{SanchezPortal97}
  
  For multiple-$\zeta$, a better scheme was proposed based on the
split-valence idea of Quantum Chemistry,\cite{Huzinaga,Poirier} but
adapted to strictly localized NAOs.\cite{Artacho99}
  In the same work, a systematic way was  proposed to generate polarization 
orbitals suited for these basis sets.
  The scheme of Ref.~\onlinecite{Artacho99} has proven to be quite efficient, 
systematic, and reasonable for a large variety of systems (for short 
reviews see Refs.~\onlinecite{Artacho99,Ordejon00}).

  In this work we go beyond previous methodologies because of two main reasons:
  $(i)$ It is always desirable to obtain the highest possible precision 
given the computational resources available, and 
  $(ii)$ it is important to know and show what is the precision attainable 
by NAO basis sets of reasonable sizes. 
  None of these points was systematically addressed in this context before.

  We explore these issues by variationally optimizing basis sets for
a variety of condensed systems. 
  The parameters defining the orbitals are allowed to vary freely to minimize 
the total energy of these systems. 
  This energy is then compared with that of converged plane-wave calculations
for exactly the same systems, including same density functional and 
pseudopotentials.
  The optimal basis sets are then tested monitoring structural and elastic
properties of the systems.

  The transferability of the basis sets optimized for particular systems
is then checked by transfering them to other systems and testing 
the same energetical, structural and elastic parameters.
  Finally, the effect of localizing the orbitals tighter than what they
variationally choose  is explored on a demanding system.

\section{Method}

  The calculations presented below were all done using density-functional
theory\cite{Hohenberg-Kohn,Kohn-Sham} (DFT) in its 
local-density\cite{Perdew-Zunger} approximation (LDA).
  Core electrons were replaced by norm-conserving 
pseudopotentials\cite{tm2} in their fully separable form.\cite{kb}
  The non-local partial-core exchange-correlation correction\cite{pcec} 
was included for Cu to improve the description of the core-valence 
interactions.

  Periodic boundary conditions were used for all systems. Molecules were
treated in a supercell scheme allowing enough empty space between molecules
to make intermolecular interactions negligible.
  For solid systems, integrations over the Brillouin zone were replaced 
by converged sums over selected $\vec k$-sets.\cite{Monkhorst-Pack}

  Thus far the approximations are exactly the same for the two different sets 
of calculations performed in this work: based on NAOs and on plane-waves (PWs).
  The calculations using NAOs were performed with the {\sc Siesta} method,
described elsewhere.\cite{SanchezPortal97,Ordejon96}
  Besides the basis itself, the only additional approximation with respect
to PWs is the replacement of some integrals in real space by sums in a 
finite 3D real-space grid, controlled by one single parameter, the
energy cutoff for the grid.\cite{Ordejon96}
  This cutoff, which refers to the fineness of the grid, was converged for
all systems studied here (200 Ry for all except for Si and H$_2$, for which
80 Ry and 100 Ry were used, respectively).
  Similarly, the PW calculations were done for converged PW 
cutoffs.\cite{cutoffs}

  Cohesive curves for the solids were obtained by fitting calculated
energy values for different unit-cell volumes to cubic, quartic and
Murnaghan-like\cite{Murnaghan} curves, a procedure giving values
to the lattice parameter, the bulk modulus and the cohesive energy
of each system.
  The bulk moduli given by the Murnaghan and quartic fits deviate from
each other by around 3\%, the Murnaghan values being the lowest
and the ones shown in the tables. The deviations between Murnaghan and 
cubic fits are of the order of 7\%.
  The other cohesive parameters do not change appreciably with the fits.

  The atomic-energy reference for the cohesive energy was taken from
the atomic calculations within the same DFT and pseudopotentials, always
converged in basis set.
  They are hence the same reference for NAOs and for PWs, the difference
in cohesive energies between the two accounting for the difference in 
the total energy of the solid.
  The isolated-atom calculations included spin polarization.

\section{Basis of numerical atomic orbitals}

  The starting point of the atomic orbitals that conform the
basis sets used here is the solution of Kohn-Sham's Hamiltonian
for the isolated pseudo-atoms, solved in a radial grid,
 with the same approximations as for the solid or molecule 
(the same exchange-correlation functional and  pseudopotential). 
  A strict localization of the basis functions is ensured either
by imposing a boundary condition, by adding a confining (divergent)
potential, or by multiplying the free-atom orbital by a cutting
function.
  We describe in the following three main features of a
basis set of atomic orbitals: size, range, and radial shape.

\subsection{Size: number of orbitals per atom}

  Following the nomenclature of Quantum Chemistry, we establish
a hierarchy of basis sets, from single-$\zeta$ to multiple-$\zeta$ 
with polarization and diffuse orbitals, covering from quick calculations
of low quality to high precision, as high as the finest obtained in
Quantum Chemistry. 
  A single-$\zeta$ (also called minimal) basis set (SZ in the following)
has one single radial function per angular momentum channel, and only for 
those angular momenta with substantial electronic population in the valence of
the free atom.
  It offers quick calculations and some insight on qualitative trends 
in the chemical bonding and other properties. 
  It remains too rigid, however, for more quantitative calculations
requiring both radial and angular flexibilization.

  Starting by the radial flexibilization of SZ, a better basis is obtained 
by adding a second function per channel: double-$\zeta$ (DZ).
  Several schemes have been proposed to generate this second function. 
  In Quantum Chemistry, the {\it split valence}\cite{Huzinaga,Szabo} scheme
is widely used: starting from the expansion in Gaussians of one atomic 
orbital, the most contracted gaussians are used to define the first
orbital of the double-$\zeta$ and the most extended ones for the second.
  Another proposal defines the second $\zeta$ as the
derivative of the first one with respect to occupation.\cite{Parrinello}
  For strictly localized functions there was a first proposal\cite{Daniel-JPCM}
 of using the excited states of the confined atoms, but it would work only 
for tight confinement.
  An extension of the split valence idea of Quantum Chemistry to strictly
localized NAOs was proposed in Ref.~\onlinecite{Artacho99} and has been used
quite succesfully in a variety of systems.
  We follow this scheme in this work, which generalizes to multiple-$\zeta$
trivially.

  Angular flexibility is obtained by adding shells of higher angular 
momentum.
  Ways to generate these so-called polarization orbitals have been
described in the literature, both for Gaussians\cite{Huzinaga,Poirier}
and for NAOs.\cite{Artacho99}
  In this work, however, they will be obtained variationally, as the rest,
within the flexibilities described below.

\subsection{Range: cutoff radii of orbitals}

  Strictly localized orbitals (zero beyond a cutoff radius) are used
in order to obtain sparse Hamiltonian and overlap matrices for linear 
scaling.
  The traditional alternative to this is based on neglecting interactions 
when they fall below a tolerance or when the atoms are beyond 
some scope of neighbours.
  Its disadvantage is that it implies a deviation
from the original Hilbert space, the total energy is no longer a 
variational magnitude, and produces numerical instabilities.
  
  For the bases made of strictly localized orbitals, the problem is finding 
a balanced and systematic way of defining all the different radii, since 
both the precision and the computational efficiency in the calculations 
depend on them.
  A scheme was proposed\cite{Artacho99} in which all radii were defined
by one single parameter, the energy shift, i.e., the energy raise
suffered by the orbital when confined.
  In this work, however, we step back from that systematic approach
and allow the cutoff radii to vary freely in the optimization
procedure (up to a maximum value of 8 a.u.).

\subsection{Shape}

  Within the pseudopotential framework it is important to keep 
the consistency between the pseudopotential and
the form of the pseudoatomic orbitals in the core region.
  This is done by using as basis orbitals the solutions
of the same pseudopotential in the free atom.
  The shape of the orbitals at larger radii depends on the
cutoff radius (see above) and on the way the localization 
is enforced.
  The first proposal\cite{Sankey} used an infinite square-well
potential.
  It has been widely and succesfully used for minimal bases within 
the ab initio tight-binding scheme of Sankey and 
collaborators\cite{Sankey} using the {\sc Fireball } program, but also 
for more flexible bases using the methodology of {\sc Siesta}.

  This scheme has the disadavantage, however, of generating 
orbitals with a discontinuous derivative at $r_c$. 
  This discontinuity is more pronounced for smaller $r_c$'s and
tends to disappear for long enough values of this cutoff.
  It does remain, however, appreciable for sensible values of
$r_c$ for those orbitals that would be very wide in the free atom.
  It is surprising how small an effect such kink produces in the
total energy of condensed systems (see below).
  It is, on the other hand, a problem for forces and stresses,
especially if they are calculated using a (coarse) finite 
three-dimensional grid.

  Another problem of this scheme is related to its defining the
basis considering the free atoms. 
  Free atoms can present extremely
extended orbitals, their extension being, besides problematic,
of no practical use for the calculation in condensed systems:
the electrons far away from the atom can be described by the
basis functions of other atoms.

  Both problems can be addressed simultaneously by adding
a soft confinement potential to the atomic Hamiltonian used
to generate the basis orbitals: it smoothens
the kink and contracts the orbital as variationally suited.
  Two soft confinement potentials have been proposed in
the literature, both of the form $V(r)=V_{\rm o} r^n$, one for $n$=2 
(Ref.~\onlinecite{Porezag}) and the other for $n$=6
(Ref.~\onlinecite{Horsfield}).
  They present their own inconveniences, however. 
  Firstly, there is no radius at which the orbitals become strictly zero, 
they have to be neglected at some point.
  Secondly, these confinement potentials affect the core
region spoiling its adaptation to the pseudopotential.
 
  This last problem affects a more traditional scheme as well, namely,
the one based on the radial scaling of the orbitals by suitable scale 
factors.
  In addition to very basic bonding arguments,\cite{Cohen-Tannoudji} it is
soundly based on restoring virial's theorem for finite bases, in the case
of coulombic potentials (all-electron calculations).\cite{Levine}
  The pseudopotentials limit its applicability, allowing only for
extremely small deviations from unity ($\sim 1\%$) in the scale factors 
obtained variationally (with the exception of hydrogen that can contract 
up to 25\%).\cite{scalef}
 
  An alternative scheme has also been proposed:\cite{Kenny}
Instead of modifying the potential, it directly modifies the
orbitals of the free atom.
  Following ideas of previous mixed-basis schemes\cite{Elsaesser}
the atomic orbital is multiplied by 
$1 - \exp [ -\alpha (r-r_c)^2 ]$ for $r < r_c$ and zero otherwise.
  This scheme does provide strict localization beyond $r_c$, but
introduces a different problem: for large $\alpha$ and small 
$r_c$ a bump appears in the orbital close to $r_c$, which becomes
a discontinuity in the wave-function in the limit of zero $\alpha$.

  In this work we propose a new soft confinement potential 
avoiding the mentioned deficiencies. 
  It is flat (zero) in the core region, starts off at
some internal radius $r_i$ with all derivatives continuous
and diverges at $r_c$ ensuring the strict localization there.
  It is
\begin{equation}
  V(r) = V_{\rm o} { e^{- { {r_c - r_i} \over {r - r_i} } } \over {r_c -r} }
\end{equation}
  In the following the different schemes are compared, their
defining parameters being allowed to change variationally.

  Finally, the shape of an orbital is also changed by the ionic character 
of the atom. 
  Orbitals in cations tend to shrink, and they swell in anions.
  Introducing a $\delta Q$ in the basis-generating free-atom calculations
gives orbitals better adapted to ionic situations in the condensed
systems.

\section{Optimization procedure}

  Given a system and a basis size, the range and shape
of the orbitals are defined as described above, depending on 
parameters.
  These are defined variationally: the energy is minimized with respect 
to all of them.
  As a robust and simple minimization method not requiring
the evaluation of derivatives, we have chosen the downhill
simplex method.\cite{Numerical}
 
  The parameters varied in an optimization are the following.
  Per atomic species there is a global $\delta Q$.
  Confinement is imposed differently for each angular momentum shell,
with its corresponding parameters that depend on the scheme used.
  Hard confinement implies one parameter per shell ($r_c$), and our
soft confinement implies three ($r_c$, $r_i$, and $V_{\rm o}$).
  One parameter ($V_{\rm o}$) is needed only in the $r^n$-confinement
schemes,\cite{Porezag,Horsfield} and two parameters in the scheme
of Kenny {\it et al.}\cite{Kenny} ($r_c$ and the width of the cutting
function).
  Finally, for each $\zeta$ beyond the first, there is a matching radius
as defined in Ref.~\onlinecite{Artacho99}.
  The values obtained for the parameters in the optimizations
described below can be obtained from the authors.

\section{Results}

\subsection{Comparison of different confinement schemes}

\begin{figure}
\epsfig{figure=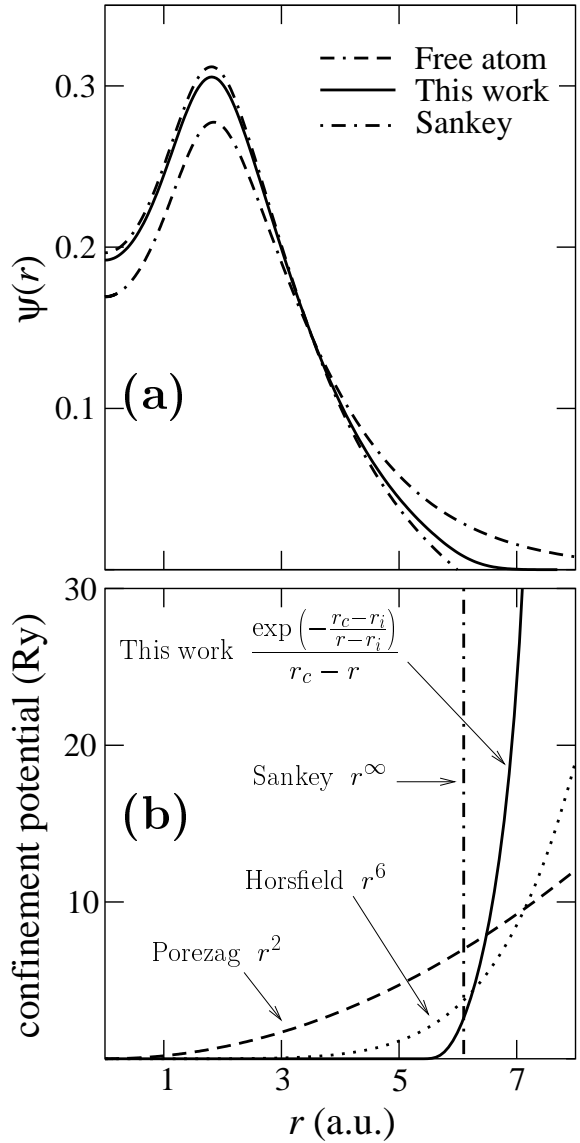,width=3in}
\caption[ ]{Shape of the 3$s$ orbital of Mg in MgO for the
different confinement schemes~(a) and corresponding potentials~(b).}
\label{MgO}
\end{figure}

  Table~\ref{schemes} shows the performance for MgO of the different
schemes described above for constructing localized atomic orbitals.
\begin{table}
\caption[ ]{Comparison of different confinement schemes
on the cohesive properties of MgO, for SZ and DZP basis sets.
The generalization of the different schemes to DZP is done
as explained in the text. Unconfined refers to using the 
unconfined pseudoatomic orbitals as basis. $a$, $B$, and $E_c$
stand for lattice parameter, bulk modulus, and cohesive energy,
respectively. The PW calculations were performed with identical
approximations as the NAO ones except for the basis. Augmented
plane-wave (LAPW) results were taken from Ref.~\onlinecite{Goniakowski}, 
and the experimental values from Ref.~\onlinecite{Finocchi}.}
\begin{tabular}{c|ccc|ccc}
& \multicolumn{3}{c|}{SZ} &
\multicolumn{3}{c}{DZP} 
\\
Basis & $a$ & $B$ & $E_{c}$ & 
   $a$ & $B$ & $E_{c}$
\\
scheme & (\AA) & (GPa) & (eV) & 
   (\AA) & (GPa) & (eV)
\\
\hline
Unconfined & 4.25 & 119 & 6.49 & & &  \\
Sankey & 4.17 & 222  & 10.89 & 4.12 & 165 & 11.82 \\
Kenny & 4.16 & 228 & 11.12 & 4.12 & 163 & 11.84 \\
Porezag & 4.18 & 196 & 11.17 & 4.09 & 183 & 11.83 \\
Horsfield & 4.15 & 221 & 11.26 & 4.11 & 168 & 11.86 \\
This work & 4.15 & 226 & 11.32 & 4.10 & 167 & 11.87 \\
\hline
PW & & & & 4.10 & 163 & 11.89 \\
LAPW & & & & 4.26 & 147 & 10.4 \\
Exp. & & & & 4.21 & 152 & 10.3 \\
\end{tabular}
\label{schemes}
\end{table}
  The basis sets of both magnesium and oxygen were variationally 
optimized for all the schemes.
  Mg was chosen because the 3$s$ orbital is very extended in 
the atom and both the kink and the confinement effects due to
other orbitals are very pronounced.
  Results are shown for a SZ (single $s$ and $p$ channels for both 
species) and a DZP basis (double $s$ and $p$ channels plus a single $d$ 
channel).
  Fig.~\ref{MgO} shows the shape of the optimal 3$s$ 
orbital for the different schemes, and the shape of the
confining potentials.

  The following conclusions can be drawn from the results:
  ($i$) Within the variational freedom
offered here, the 3$s$ orbital of Mg wants to be confined to
a radius of around 6.5 Bohr, irrespective of scheme,
which is extremely short for the free atom. 
  This confinement produces a pronounced kink in the hard scheme.
  ($ii$) The total energy is relatively insensitive to the scheme
used to generate the basis orbitals, as long as there is effective
confinement.
  ($iii$) The basis made of unconfined atomic orbitals is substantially
worse than any of the others. 
  ($iv$) The pronounced kink obtained in Sankey's hard confinement 
scheme is not substantially affecting the total energy as compared 
with the other schemes. 
  It does perturb, however, by introducing inconvenient noise in the 
energy variation with volume and other external parameters, and especially
in the derivatives of the energy.
  ($v$) The scheme proposed in this work is variationally slightly better
than the other ones, but not significantly.
  Its main advantage is the avoidance of known problems.
  In the remainder of the paper, the confinement proposed in this work
will be used unless otherwise specified.
 
\subsection{Basis convergence}

\begin{figure}
\epsfig{figure=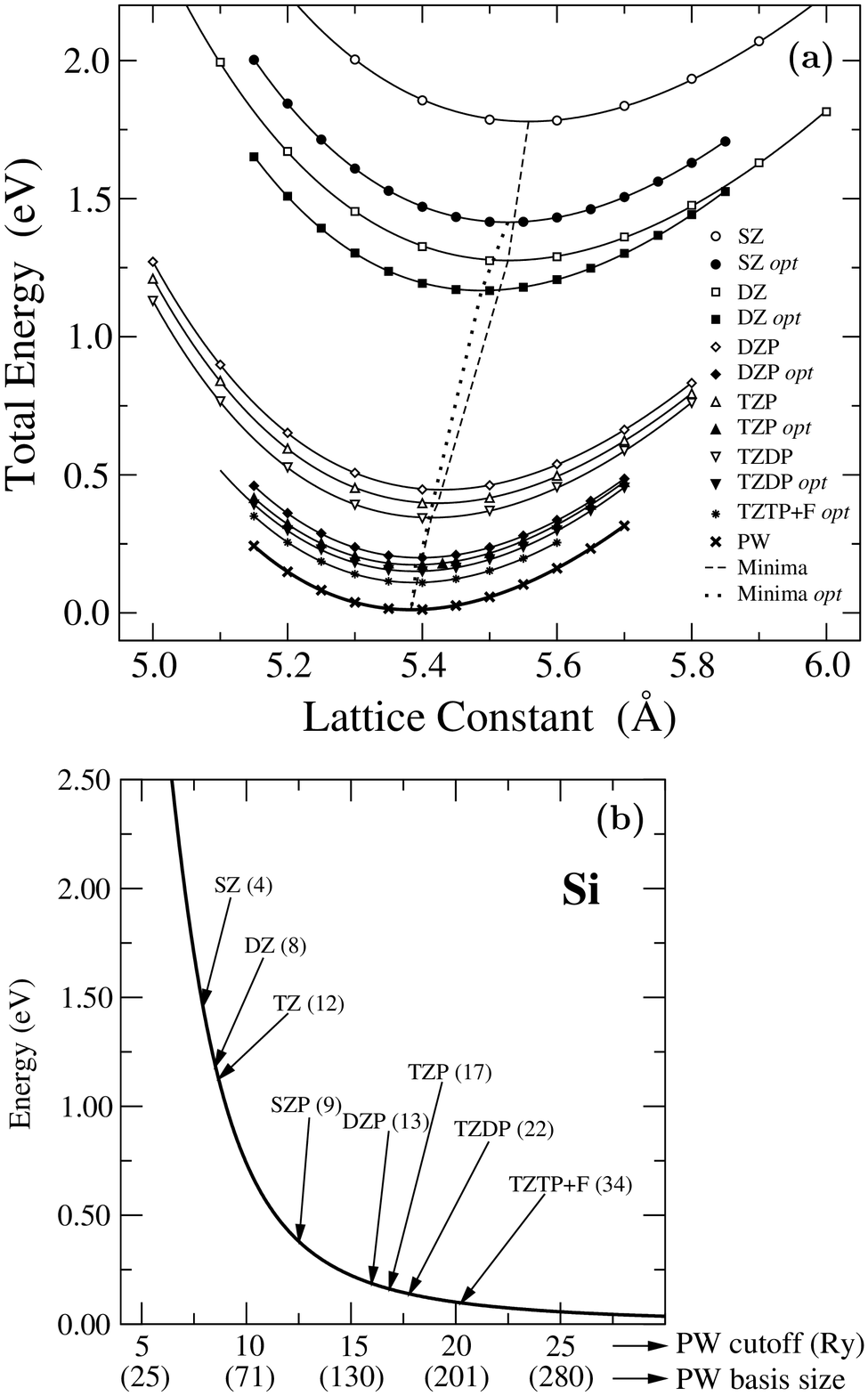,width=3in}

\caption[ ]{Convergence of NAO basis sets for bulk Si.\\ (a) Cohesive
curve for different basis sets. The lowest curve shows the PW results,
filled symbols the NAO bases of this work ($opt$), and open symbols the NAO 
bases following Ref.~\onlinecite{Artacho99}. Basis labels are like in
Table~\ref{Si-conv}, F standing for an extra $f$-shell.\\
(b) Comparison of NAO convergence with PW convergence. In parenthesis 
the number of basis functions per atom.}
\label{Si-curves}
\end{figure}

  Table \ref{Si-conv} shows how NAO bases converge for bulk silicon.
  This is done by comparing different basis sizes, each of them
optimized. 
  The results are compared to converged (50 Ry) PW results (converged
basis limit) keeping the rest of the calculation identical. 
  They are also compared to all-electron LDA results\cite{Filippi} to 
compare basis errors with the ones produced by the pseudopotentials.
  Experiment gives a reference to the accuracy of the underlying LDA.
  Fig.~\ref{Si-curves}(a) shows the cohesion curves for this system.
\begin{table} 
\caption[ ]{Basis comparisons for bulk Si. $a$, $B$, and $E_c$
stand for lattice parameter (in \AA), bulk modulus (in GPa), and 
cohesive energy (in eV), respectively. SZ, DZ, and TZ stand for 
single-$\zeta$, double-$\zeta$, and triple-$\zeta$. P stands 
for polarized, DP for doubly polarized. LAPW results were taken from 
Ref.~\onlinecite{Filippi}, and the experimental values from 
Ref.~\onlinecite{Kittel}.}
\begin{tabular}{lcccccccccc}
 & SZ & DZ & TZ & SZP & DZP & TZP & TZDP & PW & LAPW & Exp\\
\hline
$a$ & 5.52 & 5.49 & 5.48 & 5.43 & 5.40 & 5.39 & 5.39 & 5.38 & 
5.41 & 5.43 \\
$B$ & 85 & 87 & 85 & 97 & 97 & 97 & 97 & 96 & 96 & 98.8   \\
$E_{c}$ & 4.70 & 4.83 & 4.85 & 5.21 & 5.31 & 5.32 & 5.34 & 5.40 & 
5.28 & 4.63   \\
\end{tabular}
\label{Si-conv}
\end{table}

  The comparisons above are made with respect to the converged-basis 
limit, for which we used PWs up to very high cutoffs.
  It is important to distinguish this limit from the PW
calculations at lower cutoffs, as used in many computations.
  To illustrate this point, Fig.~\ref{Si-curves}(b)
compares the energy convergence for PWs and for NAOs.
  Even though the convergence of NAO results is {\it a priori} 
not systematic with the way the basis is enlarged, the sequence 
of bases presented in the Figure shows a nice convergence of total
energy with respect to basis size (the number of basis functions per 
atom are shown in parenthesis in the Figure):
  the convergence rate is similar to the one of PWs
(DZP has three times more orbitals than SZ, and a similar factor is 
found for their equivalents in PWs).
  For the particular case of Si, Fig.~\ref{Si-curves} shows that the
polarization orbitals (3$d$ shell) are very important
for convergence, more than the doubling of the basis.
  This fact is observed from the stabilization of SZP with respect
to SZ, which is much larger than for DZ.

  Fig.~\ref{Si-curves} shows that an atomic basis requires ten times 
less functions than its (energetically) equivalent PW basis, being
Si the easiest system for PWs.
  For other systems the ratio is much larger, as shown in Table~\ref{pwequi}.
\begin{table}
\caption[ ]{Equivalent PW cutoff ($E_{\rm cut}$) to optimal DZP bases 
for different systems. Comparison of number of basis functions per atom
for both bases. For the molecules, a cubic unit cell of 10 \AA\ 
of side was used.}
\begin{tabular}{cccc}
System & \# funct. DZP & \# funct. PW & $E_{\rm cut}$ (Ry)\\
\hline
$\rm H_{2}$     &  5 & 11296 & 34 \\
$\rm O_{2}$     & 13 & 45442 & 86 \\
Si              & 13 & 144   & 16 \\
Diamond         & 13 & 284   & 59 \\
$\alpha$-quartz & 13 & 923   & 76 \\
\end{tabular}
\label{pwequi}
\end{table}

  It is important to stress that deviations smaller than
the ones due to the pseudopotential or the DFT used are obtained
with a relatively modest basis size as DZP.
  This fact is clear in Table~\ref{Si-conv} for Si, and in 
Table~\ref{solids} for other systems.
\begin{table} 
\caption[ ]{Basis comparisons for different solids. $a$, $B$, and $E_c$
stand for lattice parameter (in \AA), bulk modulus (in GPa), and
cohesive energy (in eV), respectively.}
\begin{tabular}{llccccc}
& & Exp & LAPW & Other PW & PW & DZP \\
\hline
    Au 
    & $a$
    & 4.08\tablenotemark[1]
    & 4.05\tablenotemark[2]
    & 4.07\tablenotemark[9]
    & 4.05 
    & 4.07   \\

    & $B$
    & 173\tablenotemark[1]
    & 198\tablenotemark[2]
    & 190\tablenotemark[9]
    & 195 
    & 188 \\

    & $E_{c}$
    & 3.81\tablenotemark[1]
    & -
    & -
    & 4.36 
    & 4.13 \\
\hline
    MgO & $a$
    & 4.21\tablenotemark[3]
    & 4.26\tablenotemark[4]
    & -
    & 4.10  
    & 4.11  \\

    & $B$
    & 152\tablenotemark[3]
    & 147\tablenotemark[4]
    & -
    & 164 
    & 167 \\

    & $E_{c}$ 
    & 10.30\tablenotemark[3]
    & 10.40\tablenotemark[4]
    & -
    & 11.89 
    & 11.87 \\
\hline
    C  
    & $a$
    & 3.57\tablenotemark[1]
    & 3.54\tablenotemark[6]
    & 3.54\tablenotemark[5]
    & 3.53 
    & 3.54 \\

    & $B$ 
    & 442\tablenotemark[1]
    & 470\tablenotemark[6]
    & 436\tablenotemark[5]
    & 459
    & 453 \\

    & $E_{c}$ 
    & 7.37\tablenotemark[1]
    & 10.13\tablenotemark[6]
    & 8.96\tablenotemark[5]
    & 8.89
    & 8.81  \\
\hline
    Si 
    & $a$
    & 5.43\tablenotemark[1]
    & 5.41\tablenotemark[7] 
    & 5.38\tablenotemark[5]
    & 5.38
    & 5.40  \\

    & $B$ 
    & 99\tablenotemark[1]
    & 96\tablenotemark[7] 
    & 94\tablenotemark[5]
    & 96
    & 97   \\

    & $E_{c}$ 
    & 4.63\tablenotemark[1]
    & 5.28\tablenotemark[7] 
    & 5.34\tablenotemark[5]
    & 5.40
    & 5.31  \\
\hline
    Na 
    & $a$
    & 4.23\tablenotemark[1] 
    & 4.05\tablenotemark[11]
    & 3.98\tablenotemark[5]
    & 3.95
    & 3.98  \\

    & $B$ 
    & 6.9\tablenotemark[1] 
    & 9.2\tablenotemark[11]
    & 8.7\tablenotemark[5]
    & 8.8
    & 9.2  \\

    & $E_{c}$ 
    & 1.11\tablenotemark[1] 
    & 1.44\tablenotemark[8] 
    & 1.28\tablenotemark[5]
    & 1.22
    & 1.22  \\
\hline
    Cu 
    & $a$
    & 3.60\tablenotemark[1] 
    & 3.52\tablenotemark[2] 
    & 3.56\tablenotemark[5]
    & -
    & 3.57  \\

    & $B$ 
    & 138\tablenotemark[1] 
    & 192\tablenotemark[2] 
    & 172\tablenotemark[5]
    & -
    & 165  \\

    & $E_{c}$ 
    & 3.50\tablenotemark[1] 
    & 4.29\tablenotemark[10] 
    & 4.24\tablenotemark[5]
    & -
    & 4.37  \\
\hline
    Pb 
    & $a$
    & 4.95\tablenotemark[1] 
    & -
    & 4.88 
    & -
    & 4.88  \\

    & $B$ 
    & 43\tablenotemark[1] 
    & -   
    & 54
    & -
    & 64 \\

    & $E_{c}$ 
    & 2.04\tablenotemark[1] 
    & -
    & 3.77 
    & -
    & 3.51  
\end{tabular}
\tablenotetext[1]{C. Kittel, Ref. \onlinecite{Kittel}}
\tablenotetext[2]{A. Khein, D. J. Singh, C.J. Umrigar, Ref. \onlinecite{Khein} }
\tablenotetext[3]{F. Finocchi, J. Goniakowski, C. Noguera, Ref. \onlinecite{Finocchi}}
\tablenotetext[4]{J. Goniakowski, C. Noguera, Ref. \onlinecite{Goniakowski} }
\tablenotetext[5]{M. Fuchs, M. Bockstedte, E. Pehlke, M. Scheffler, 
                  Ref. \onlinecite{Fuchs}}
\tablenotetext[6]{N. A. W. Holzwarth et al, Ref. \onlinecite{Holzwarth} }
\tablenotetext[7]{C. Filippi, D. J. Singh, and C. J. Umrigar, 
                  Ref.\onlinecite{Filippi}}
\tablenotetext[8]{M. Sigalas {\it et al.}, Ref. \onlinecite{Sigalas} }
\tablenotetext[9]{B. D. Yu and M. Scheffler, Ref. \onlinecite{Yu}}
\tablenotetext[10]{P. H. T. Philipsen and E. J. Baerends, Ref. \onlinecite{Philipsen}}
\tablenotetext[11]{J. P. Perdew {\it et al.}, Ref. \onlinecite{Perdew}}
\label{solids}
\end{table}
  Table~\ref{solids} summarizes the cohesion results for
a variety of solids of different chemical kind.
  They are obtained with optimal DZP basis sets.
  It can be observed that DZP offers results in good agreement
with converged-basis numbers, showing the convergence of 
properties other than the total energy.
  The deviations are similar or smaller than those introduced by 
LDA or by the pseudopotential.\cite{Fuchs}

\section{Transferability}

  To what extent do optimal bases keep their performance
when transfered to different systems than the ones they
were optimized for?
  This is an important question, since if the performance
does not suffer significantly, one can hope to tabulate basis sets per
species, to be used for whatever system.
  If the transferability is not satisfactory, a new basis set should then
be obtained variationally for each system to be studied.
  Of course the transferability increases with basis size, since the 
basis has more flexibility to adapt to different environments.
  A thorough study of transferability, therefore, implies many calculations
in different systems, for different bases and different basis sizes.
  In this work we limit ourselves to try it on DZP
bases for a few representative systems.

  Satisfactory transferability has been obtained when checking in
MgO the basis set optimized for Mg bulk and O in a water molecule.
  Similarly the basis for O has been tested in H$_2$O and O$_2$,
and the basis for C in graphite and diamond.
  Again, the results show deviations due to the basis that are
smaller than the errors introduced by the pseudopotentials and/or
the DFT functional.
  The results are shown in Table~\ref{transfer}.
\begin{table}
\caption[ ]{Transferability of basis sets.
``Transf" stands for the DZP basis transfered from other systems, while
``Opt" refers to the DZP basis optimized for the particular system.
For MgO the basis was transfered from bulk Mg and an H$_2$O molecule,
for graphite the basis was transfered from diamond, and for H$_2$O it
was taken from H$_2$ and O$_2$.
$a$, $B$, and $E_c$ stand for lattice parameter, bulk modulus, and
cohesive energy, respectively. $\Delta E$ stands for the energy 
difference per atom between graphite and a graphene plane.
$E_b$ is the binding energy of the molecule.}
\begin{tabular}{cc|ccc}
 System & Basis &
 \multicolumn{3}{c}{Properties} 
 \\
 \hline
 MgO &        & $a$ (\AA) & $B$ (GPa) & $E_{c}$ (eV) \\
     & Transf & 4.13 & 157 & 11.81 \\
     & Opt    & 4.10 & 167 & 11.87 \\
     & PW     & 4.10 & 164 & 11.89 \\
     & LAPW   & 4.26 & 147 & 10.40 \\
     & Exp    & 4.21 & 152 & 10.30 \\
 \hline
 Graphite & & $a$ (\AA) & $c$ (\AA) & $\Delta E$ (meV) \\
          & Transf & 2.456 & 6.50 & 38 \\
          & PW\tablenotemark[1] & 2.457 & 6.72 & 24 \\
          & Exp\tablenotemark[2]& 2.456 & 6.674 & 23\tablenotemark[3] \\
 \hline
 H$_2$O   & & $d_{\rm O-H}$ (\AA) & $\theta_{\rm O-H-O}$ (deg) & $E_b$ (eV)  \\
          & Transf & 0.975 & 105.0 & 12.73 \\ 
          & Opt & 0.972 & 104.5 & 12.94 \\ 
          & PW & 0.967 & 105.1 & 13.10 \\ 
          & LAPW\tablenotemark[4] & 0.968 & 103.9 & 11.05\\ 
          & Exp\tablenotemark[5] & 0.958 & 104.5 & 10.08 \\
\end{tabular}
\tablenotetext[1]{M. C. Schabel, Ref. \cite{Schabel}}
\tablenotetext[2]{Y. Baskin, Ref. \cite{Baskin}}
\tablenotetext[3]{L. A. Girifalco, Ref. \cite{Girifalco}}
\tablenotetext[4]{P. Serena, Ref. \cite{Serena}}
\tablenotetext[5]{G. Herzberg, Ref. \cite{Herzberg}}
\label{transfer}
\end{table}

  Table~\ref{quartz} shows the results for the structural parameters
of SiO$_2$ in its $\alpha$-quartz structure.
\begin{table} 
\caption[ ]{Performance of the basis of Si and O as optimized in c-Si and
in a water molecule, respectively, for the structural parameters of $\alpha$-quartz.}
\begin{tabular}{lcccccc}
& Exp \tablenotemark[1]& PW\tablenotemark[2] & PW\tablenotemark[3] & 
PW\tablenotemark[4] & PW\tablenotemark[5] & DZP \\
\hline
    $a$ (\AA) & 4.92 & 4.84 & 4.89 & 4.81 & 4.88 & 4.85 \\
    $c$ (\AA) & 5.41 & 5.41 & 5.38 & 5.32 & 5.40 & 5.38 \\
    $d_{\rm Si-O}^1$ (\AA) & 1.605 & 1.611 & 1.60 & 1.605 & - & 1.611 \\
    $d_{\rm Si-O}^2$ (\AA) & 1.614 & 1.617 & 1.60 & 1.605 & - & 1.611 \\
    $\alpha_{\rm Si-O-Si}$ (deg) & 143.7 & 140.2 & - & 139.0 & - & 140.0 \\
\end{tabular}
\tablenotetext[1]{L. Levien, C. T. Prewitt and D. J. Weidner, Ref.~\onlinecite{Levien}.}
\tablenotetext[2]{P. Sautet (unpublished) using ultrasoft pseudopotentials.\cite{vasp}}
\tablenotetext[3]{D. R. Hamann, Ref.~\onlinecite{Hamann}.}
\tablenotetext[4]{G.-M.Rignanese {\it et al.}, Ref.~\onlinecite{Car}.}
\tablenotetext[5]{F. Liu {\it et al.}, Ref.~\onlinecite{Liu}, using ultrasoft 
pseudopotentials.}
\label{quartz}
\end{table}
  The DZP numbers have been obtained for a basis that was 
optimized not for $\alpha$-quartz itself, but for bulk silicon 
for the Si basis and for the water molecule for the O basis. 
  SiO$_2$ was chosen because its being quite sensitive to
many approximations and in particular to the basis set.
  It was hard to converge for previous NAO schemes,\cite{Artacho99}
giving\cite{Sautet} typically longer Si-O bonds (with deviations of 
around 1.5\%) and smaller unit cells (deviations of around 1.5\% 
and 2\% for the $a$ and $c$ parameters, respectively).
  The results of Table ~\ref{quartz} are very satisfactory, showing 
($i$) the good performance of NAOs, ($ii$) their transferability in 
this case, and ($iii$) the improvement of the basis sets proposed 
here over previous bases.

\section{Limiting the range}

  In this work we have concentrated on variationally optimized
basis sets, allowing the cutoff radii for the different orbitals
to vary freely, as long as the orbitals remained strictly localized.
  This was done in the spirit of exploring the capabilities of the
NAO basis sets.
  Some orbitals demanded reasonably short values of $r_c$, others
chose long ranges.
  As mentioned earlier, the range of the orbitals is important for
the efficiency in the calculations.
  Therefore, further work will be very important to explore 
the possibility of enforcing smaller ranges in reasonably balanced 
ways and its effect on the precision.
  A systematic study in this direction will be subject of future
work, we have limited ourselves here to illustrate the nature of
the problem in the particular example of $\alpha$-quartz.

  The basis has been optimized as before (Si in bulk Si and O in
H$_2$O), but imposing now tighter $r_c$'s. 
  The results are summarized in Table~\ref{quartzshort}.
\begin{table} 
\caption[ ]{Tightening the confinement of the basis in $\alpha$-quartz.}
\begin{tabular}{ccc|ccc|ccccc}
\multicolumn{3}{c|}{$r_c^{\rm Si}$ (a.u.)} &
\multicolumn{3}{c|}{$r_c^{\rm O}$ (a.u.)} &
$a$ & $c$ & $d_{\rm Si-O}^1$ & $d_{\rm Si-O}^2$ &
$\alpha_{\rm Si-O-Si}$
\\
$s$ & $p$ & $d$ & $s$ & $p$ & $d$ &
(\AA) & (\AA) & (\AA) & (\AA) & (deg)   \\
\hline
 8.0&8.0&8.0& 8.0 & 8.0 & 8.0 & 4.85 & 5.38 & 1.611 & 1.612 &140.0 \\
 6.0&6.0&6.0& 8.0 & 8.0 & 8.0 & 4.85 & 5.35 & 1.607 & 1.608 &140.0 \\
 6.0&6.0&6.0& 5.0 & 5.0 & 5.0 & 4.74 & 5.29 & 1.610 & 1.610 &134.0 \\
 6.0&6.0&6.0& 4.5 & 4.5 & 4.5 & 4.69 & 5.26 & 1.610 & 1.610 &132.0 \\
 6.0&6.0&6.0& 5.0 & 6.5 & 4.0 & 4.84 & 5.36 & 1.607 & 1.608 &139.7 \\
 5.6&6.3&4.2& 4.0 & 5.3 & 2.8 & 4.81 & 5.34 & 1.607 & 1.610 &138.2 \\
\end{tabular}
\label{quartzshort}
\end{table}
  The constraining of Si orbitals to 6.0 a.u. affects the geometry only 
slightly, whilst the contraction of the O orbitals to 5.0 a.u. 
implies a substantial contraction of the cell due to the decrease of
Si-O-Si angle rather than the shortening of the Si-O bond. 
  Note that, from an atomic perspective, the confinement of Si to
6 a.u. is tighter than the 5 a.u. confinement of O.

  Allowing for different $r_c$'s for the different channels we observe
that the shrinking of the cell is avoided keeping a long $p$
orbital for O, the $s$ and $d$ remaining comparably shorter, indicating
the (expected\cite{Artacho99}) different ``compressibilities" of
the different orbitals.
  As a candidate of unifying criterion, we tested the simple definition
of $r_c$ for each channel as the value of the radius at which the 
unconstrained orbital would have a defined value (0.01 a.u.$^{-3}$). 
  Even though the energy raise was appreciable (about 70 meV per atom), 
the geometry retained an acceptable precision, the orbitals being quite 
short, thus allowing quite efficient calculations.
  Further work is, however, needed to explore in detail this and other
possibilities.

\section{Conclusions}
 
  The variational optimization of NAO basis sets for different systems
allows us to draw the following conclusions. 
  ($i$) The performance of NAO basis sets of modest size as DZP 
is very satisfactory for the systems tried. 
  For this basis size, the deviation from basis convergence is smaller than
errors due to the pseudopotential or DFT used.
  ($ii$) The bases obtained here represent a substantial improvement over 
previous NAO basis sets. In particular, the optimization in condensed 
systems offers better and more efficient bases than purely atomic schemes.
  ($iii$) The radial shapes of the orbitals obtained as proposed in this work
offer better bases than previous schemes from a variational
point of view, albeit not a substantial difference is obtained.
  ($iv$) The elimination of the discontinuity in the derivative, while 
retaining strict localization and leaving the core region untouched, 
gives bases of better quality from the point of view of the energy,
its derivatives, and computational efficiency.
  ($v$) The bases obtained showed enough transferability to expect
that a basis tabulation would be useful, and that the optimization
of the basis for each particular system will not be necessary.
  Finally, the selective sensitivity to orbital-range tightening has 
been shown, making clear the need of further work systematically to
control the cutoff radii for improving efficiency without loss of 
precision.

\acknowledgments

  We acknowledge discussions and input from J. M. Soler, E. Anglada,
Z. Barandiar\'an, L. Seijo, P. Ordej\'on, and A. Garc\'{\i}a. 
  This work was supported by the Spanish Direcci\'on General de 
Investigaci\'on under grant BFM2000-1312 and by the Fundaci\'on Ram\'on 
Areces of Spain.


\begin{references}
\bibitem{cutoffs}
  A plane wave cutoff of 100 Ry was used for MgO and diamond; 
  62 Ry for Au; 50 Ry for Si and Na; 90 Ry for
  $\alpha$-quartz and 140 Ry for H$_2$O.
\bibitem{scalef}
  In the context of NAOs it has been used for SZ bases. In liquid silicon 
  simulations\cite{Gabriel} variational contractions of 2\% for the 3$p$ 
  and 0\% for the 3$s$ orbitals of Si were used.
  For carbon in fullerenes\cite{Emilio} the contraction was of 5\% for 
  2$s$ and 2$p$. And the 1$s$ orbital of hydrogen contracted\cite{Emilio}
  20\% in H$_2$ and 25\% in H$_2$O.
\bibitem{web}
  Values for the parameters can be obtained upon request.
\end{references}
\end{document}